\documentclass[aip,twocolumn,apl,reprint,amsmath,amssymb,superscriptaddress]{revtex4-1}

\usepackage{graphicx}
\usepackage{dcolumn}
\usepackage{bm}

\begin{document}

\title{Electrometry using the quantum Hall effect in a bilayer 2D electron system }

\author{L.H. Ho}
\email{laphang@phys.unsw.edu.au}

\author{L.J. Taskinen}
\author{A.P. Micolich}
\author{A.R. Hamilton}
\email{Alex.Hamilton@unsw.edu.au}
\affiliation{School of Physics,
University of New South Wales, Sydney NSW 2052, Australia}

\author{P. Atkinson}
\author{D. A. Ritchie}
\affiliation{Cavendish Laboratory, University of Cambridge, Cambridge CB3 0HE, United Kingdom}

\date{\today}

\begin{abstract}
We discuss the development of a sensitive electrometer that utilizes
a two-dimensional electron gas (2DEG) in the quantum Hall regime. As
a demonstration, we measure the evolution of the Landau levels in a
second, nearby 2DEG as the applied perpendicular magnetic field is
changed, and extract an effective mass for electrons in GaAs that
agrees within experimental error with previous measurements.
\end{abstract}

\maketitle

The integer~\cite{vonKlitzingPRL80} and fractional~\cite{TsuiPRL82}
quantum Hall effects are two of the most significant discoveries to
emerge from several decades of intense study of two dimensional
electron systems (2DESs). The density of states, which is central to
understanding the physics of the quantum Hall effect, is not easily
accessible via traditional transport measurements alone. Instead,
the density of states is usually accessed via measurements of
thermodynamic quantities such as the specific
heat,~\cite{GornikPRL85} magnetization,~\cite{EisensteinPRL85} or
compressibility.~\cite{EisensteinPRB94} Studies of the magnetization
of 2DESs in the quantum Hall regime have been particularly fruitful
but at the same time extremely difficult.~\cite{UsherJPCM09} Another
way to extend beyond transport studies is to measure the chemical
potential directly using single-electron transistor (SET)
electrometers located on the heterostructure
surface.~\cite{HuelsPRB04} Although electrometry is considerably
easier than magnetometry, SETs can be difficult to fabricate and are
very sensitive to local fluctuations, causing significant
measurement noise. A variant of this approach was pioneered by
Kawano and Okamoto, who created a scanning electrometer using a
quantum Hall effect device,~\cite{KawanoAPL04} and used it to study
Landau level scattering in a second 2DES.~\cite{KawanoPRB04}

In this work, we take these earlier electrometry measurements one
step further to produce a sensitive electrometry system for studying
a 2DES in the quantum Hall regime. Our electrometer uses the close
proximity and strong capacitive coupling between two 2DESs in a
double quantum well heterostructure -- using one 2DES as a quantum
Hall effect electrometer for the other. Our design is far simpler to
implement than the previous designs by Huels {\it et al.} and Kawano
and Okamoto. Additionally, the closer proximity and larger interface
area should result in increased sensitivity and reduced noise
compared to earlier implementations.~\cite{KawanoAPL04,KawanoPRB04}
As a demonstration of our device, we use it to map the evolution of
the Landau levels (LLs) in a 2DES as a function of applied magnetic
field. This technique could be used to investigate the effective
mass $m^{*}$ and the Lande $g$-factor of 2D electron systems in less
studied semiconductor heterostructures such as InGaAs/InP.

The device is fabricated on an AlGaAs/GaAs heterostructure (A2264)
featuring two $20$~nm wide GaAs wells separated by a $30$~nm AlGaAs
barrier. This gives an effective 2DES separation $d = 50$~nm.
Figures 1(a) and (b) show a top-view micrograph and a side-view
schematic of the device, which is etched into a Hall bar
configuration with NiGeAu ohmic contacts that penetrate both quantum
wells. The device has five gates: a top-gate (shaded green) biased
at $V_{\rm TG}$ that controls the electron density in the upper
2DES, and a set of four depletion gates (shaded white) to sever the
connection between the upper 2DES and the ohmic contacts at the ends
and sides of the Hall bar. All electrical measurements were
performed at $\sim 50$~mK using four-terminal lock-in techniques
with an excitation voltage of $100~\mu$V at $17$~Hz.
Characterization of the device revealed that the upper (lower) 2DES
has a mobility of $1.2 \times 10^{6}$~cm$^{2}/$Vs ($1.4 \times
10^{6}$~cm$^{2}/$Vs) and density $n_{\rm T} = 2.00 \times
10^{11}$~cm$^{-2}$ ($n_{\rm B} = 1.98 \times 10^{11}$~cm$^{-2}$)
with the top-gate unbiased.

\begin{figure}
\includegraphics[width=8cm]{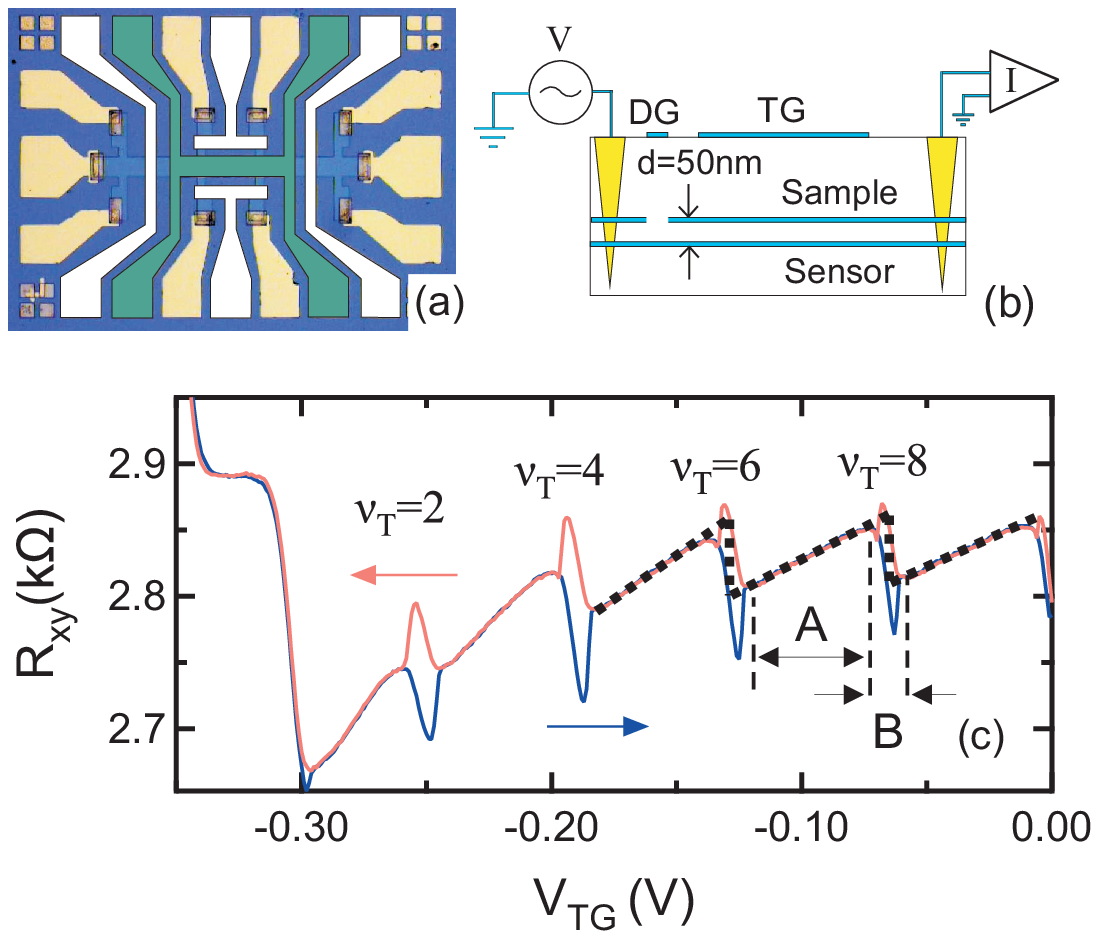}
\caption{\label{fig:schematic}} (a) An optical micrograph of the
device (b) A schematic of the device and measurement circuit. (c)
The Hall resistance $R_{\rm xy}$ of the sensor 2DES (solid lines) vs
top gate voltage $V_{\rm TG}$ at $B = 0.924$~T for a sweep from
$V_{\rm TG} = -0.35$~V to $0$~V (blue solid line) and back to
$-0.35$~V (red solid line). The dotted black line is a guide to the
eye for the ideal equilibrium behavior of $R_{\rm xy}$ to highlight
the sharp jumps at even filling factors $\nu_{\rm T}$. The regions A 
and B correspond to those in Fig.~2(a)
\end{figure}

We now discuss the operating concept for our device. The lower 2DES
is used as a quantum Hall effect electrometer (hereafter referred to
as the `sensor') for the upper 2DES (hereafter `sample'). During the
measurement, three of the four depletion gates are biased. This
isolates the sample from the measurement circuit, aside from a
connection to ground via the drain contact, which allows the density
in the sample to respond to changes in $V_{\rm TG}$. Current thus
passes only through the sensor 2DES, and we measure Hall resistance
$R_{\rm xy}$ of that 2DES as our sensor output (the longitudinal
resistance $R_{\rm xx}$ is measured simultaneously). It is important
to note that $R_{\rm xy}$ is sensitive to both electric and magnetic
fields, and can thus detect three distinct events: changes in the
magnetic field $B$, changes in the sensor density due to changes in
$V_{\rm TG}$ if the upper 2DES is depleted, and changes in the
sensor density due to changes in the chemical potential $\Delta
\mu_{\rm T}$ in the sample 2DES. The first allows us to set an
operating point for the sensor, the second allows us to characterize
the sensor, and the latter is the quantity we seek to measure. The
coupling between the two 2DESs is capacitive, yielding a change in
sensor density:

\begin{align}
\label{eqn1} \Delta n_{\rm B} = \frac{\epsilon}{e^{2}d} \Delta \mu_{\rm T}
\end{align}
in response to a change in $\Delta \mu_{\rm T}$. The dielectric
constant $\epsilon$ for the AlGaAs barrier between 2DESs can be
directly measured, and is obtained from a comparison of the slopes of
$n_{\rm T}$ vs $V_{\rm TG}$ when sample 2DES is populated, and
$n_{\rm B}$ vs $V_{\rm TG}$ when then sample 2DES is depleted using
a modified parallel-plate capacitor model. Low field measurements of
$R_{\rm xy}$ are used to obtain $n_{\rm T}$ and $n_{\rm B}$, and we
obtain $\epsilon = 10.2 \epsilon_{\rm 0}$. When mapping the Landau
levels in the sample 2DES, the sensor 2DES will also be in the
quantum Hall regime, resulting in maximum sensitivity in the middle
of a quantum Hall transition where $R_{\rm xy}$ changes rapidly, and
zero sensitivity in the quantum Hall plateau. Although this limits
the operating range, an operating point that gives good sensitivity
is easily established.

We now show a typical measurement obtained with our device. The
chosen operating field $B = 0.924$~T corresponds to a sensor 2DES
filling factor $\nu_{\rm B} \approx 9.5$. In Fig. 1(c) we plot the
sensor output $R_{\rm xy}$ against $V_{\rm TG}$ starting at $V_{\rm
TG} = -0.35$~V and increasing to $0$~V (blue solid line), and then
returning to $-0.35$~V (red solid line). For $V_{\rm TG} < V_{\rm
depl} \sim -0.3$~V the sample 2DES is fully depleted and the
top-gate acts directly on the sensor 2DES. In this region, decreases
in $V_{\rm TG}$ reduce $n_{\rm B}$ and lead to a rising $R_{\rm
xy}$. In contrast, when $V_{\rm TG} > V_{\rm depl}$ the sample 2DES
is populated and changes in $R_{\rm xy}$ directly reflect changes in
the sample 2DES chemical potential $\mu_{\rm T}$ via Eq.~1. There
are two key features for the data in Fig. 1(c) for $V_{\rm TG} >
V_{\rm depl}$. Firstly, the red and blue traces separate markedly
when the sample 2DES filling factor $\nu_{\rm T}$ takes an even
integer value. This hysteresis is due to non-equilibrium currents in
the edge states of the sample 2DES, and will be discussed in detail
elsewhere.

\begin{figure}
\includegraphics[width=8cm]{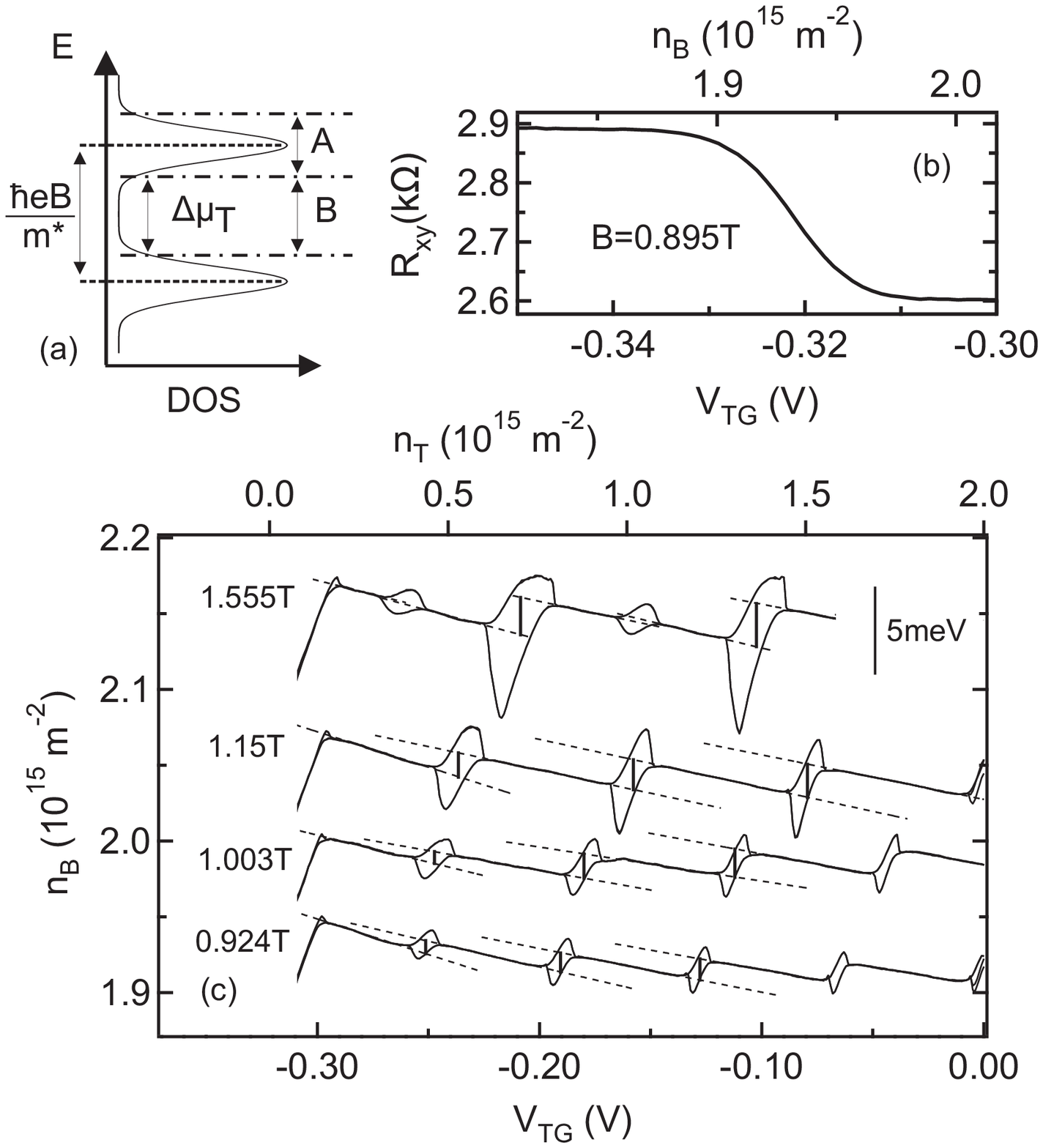}
\caption{\label{fig:sweeprate}} (a) A schematic of the density of
states (DOS) of the sample 2DES in the quantum Hall regime. (b)
$R_{\rm xy}$ vs $V_{\rm TG}$ (bottom axis) and the corresponding
$n_{\rm B}$ (top axis) at $B = 0.895$~T showing the transition
between $\nu = 9$ (right) and $10$ (left). (c) The sensor 2DES
density $n_{\rm B}$ vs $V_{\rm TG}$ (bottom axis) and the
corresponding $n_{\rm T}$ (top axis) at four different operating
points $B = 0.924$, $1.003$, $1.15$ and $1.555$~T. The latter three
traces have been vertically offset by $3$, $8$ and $16 \times
10^{13}$~m$^{-2}$ for clarity. The equivalent scale in chemical
potential is shown by the scale bar (upper right corner). The dashed
black lines show the straight line fits to the non-hysteretic
regions while the vertical solid lines at each hystersis `loop'
indicate the corresponding $\Delta \mu_{T}$ plotted in Fig.~3.
\end{figure}

The second is the sawtooth structure in the measured $R_{\rm xy}$
when the sample 2DES is populated, as highlighted by the black
dotted line at $V_{\rm TG} > -0.18$~V in Fig.~1(c). There are two
mechanisms contributing to this structure -- the periodic modulation
of the density of states in the quantum Hall regime (see Fig.~2(a))
and the well-known negative compressibility effect observed in
bilayer 2D systems.~\cite{EisensteinPRL92,MillardAPL96} Starting in
region A in Fig.~2(a), the chemical potential $\mu_{\rm T}$
coincides with a Landau level where the density of states (DOS) is
large. Here small changes in $n_{\rm T}$ produce only small changes
in $\mu_{\rm T}$, and this {\it should} produce a gentle increase in
$R_{\rm xy}$ as $V_{\rm TG}$ is increased in the corresponding
region in Fig.~1(c). Instead, we observe a gentle decrease in
$R_{\rm xy}$ caused by negative compressibility, and this is
consistent with earlier studies of bilayer 2D systems in the quantum
Hall regime.~\cite{EisensteinPRL92} Eventually we reach region B; here
the DOS is very small, and small changes in $V_{\rm TG}$ produce a
very rapid rise in $\mu_{\rm T}$. This rise overwhelms the negative
compressibility to produce a corresponding sudden drop in $\Delta
R_{\rm xy}$, as shown in Fig.~1(c).

Extracting $\Delta \mu_{\rm T}$ from the measured $\Delta R_{\rm
xy}$ vs $V_{\rm TG}$ data involves two steps. First, we need to
characterize the sensor 2DES to relate $\Delta R_{\rm xy}$ to
$\Delta n_{\rm B}$, then we can simply use Eq.~1 to convert $\Delta
n_{\rm B}$ to $\Delta \mu_{\rm T}$. The sensor characterization
involves measuring $R_{\rm xy}$ versus $V_{\rm TG}$ with the sample
2DES depleted. This initially appears straightforward, but is more
complicated because we set the operating point as the middle of a
quantum Hall transition to maximize the sensitivity. To extract
$\Delta \mu_{\rm T}$, we need to map $\Delta R_{\rm xy}$ to $\Delta
n_{\rm B}$ for the whole transition, but as Fig.~1(c) highlights,
only half of the transition is accessible if this is done at the
operating point. We overcome this by performing the characterization
at a slightly lower field $B = 0.895$~T. As Fig. 2(b) shows, this
allows us to map the entire transition without repopulating the
sample 2DES. The $\sim 3\%$ difference between the operating point
$B = 0.924$~T and the characterization point $B = 0.895$~T has a
negligible effect on the resulting calibration. The $n_{\rm B}$
versus $V_{\rm TG}$ that results from applying the calibration in
Fig.~2(b) to the data in Fig.~1(c) is shown as the bottom trace in
Fig.~2(c). Finally, the relationship between $n_{\rm T}$ (top axis)
and $V_{\rm TG}$ (bottom axis) in Fig.~2(b) is obtained from low
field Hall measurements with $\Delta n_{\rm B} = \Delta V_{\rm TG}
\times (4.01 \times 10^{15} m^{-2}/V)$.

\begin{figure}
\includegraphics[width=8cm]{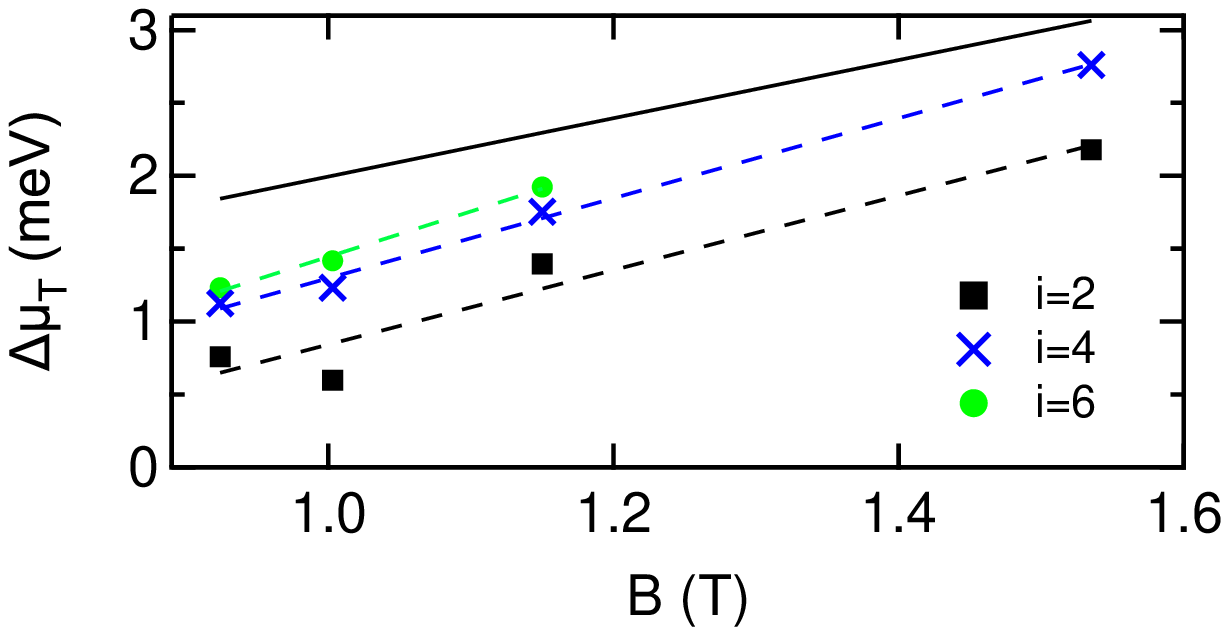}
\caption{\label{fig:fandiagram}} The energy spacing $\Delta\mu_{\rm
T}$ between adjacent Landau levels as a function of $B$. The
transitions occur at the points where the filling factor $\nu_{\rm
T}$ is an even integer $i$. The spacings have been measured for $i =
2$, $4$ and $6$. The dashed lines are a straight line fit for each
$i$, whereas the solid line shows the expected spacing $\hbar
eB/m^{*}$.
\end{figure}

We now focus on using our device to map the evolution of the three
lowest spin-degenerate Landau levels $i = 2$, $4$, and $6$ with $B$
for the sample 2DES, as shown in Fig.~3. In addition to the data for
$B = 0.924$~T discussed above, we obtain $n_{\rm B}$ vs $V_{\rm TG}$
data for three other operating points $B = 1.003$, $1.15$, and
$1.555$~T, each with its own sensor calibration performed at an
appropriate nearby $B$. These three additional traces are presented
in Fig.~2(c) and have been vertically offset for clarity. In
obtaining the Landau level spacings $\Delta \mu_{\rm T}$ for Fig.~4,
we need to overcome the obscuring effect of the hysteresis, and even
in its absence, account for the contribution to $n_{\rm B}$ versus
$V_{\rm TG}$ from the negative compressibility. We have devised a
simple method for doing this which involves three steps.
First we take linear fits to the sloped regions either side of the
LL transition, shown by the dashed lines in Fig.~2(c). We
then measure the vertical distance $\Delta n_{\rm B}$ between these
two extrapolated fits at the LL transition point, as shown by the
short vertical lines in Fig.~2(c). The transition point is assigned
as the average of the position of the extrema in the up and down
sweeps of the hysteresis loops. 
Finally, we obtain the corresponding chemical potential
change $\Delta \mu_{\rm T}$ using Eq.~1. The use of the fits to the
sloped regions adjacent to the transition in this process
automatically corrects for the negative compressibility.

The extracted $\Delta \mu_{\rm T}$ values are plotted versus $B$ in
Fig.~3. The solid line indicates the expected value of the LL
spacing $\hbar eB/m^{*}$. We have used $m^{*} = 0.058 m_{\rm e}$
rather than the more typical value $m^{*} = 0.067 m_{\rm e}$ to
account for the reduced effective mass at the low densities used in
our experiment.~\cite{ColeridgeSS96} The data for each LL follows a
linear trend as indicated by the dashed lines in Fig.~3, however in
each case, they sit well below the expected value (solid line). We
attribute this discrepancy to Landau level
broadening.~\cite{EisensteinPRL85} This broadening is partly due to
disorder, and increases as the mobility is lowered. A well known
property of modulation doped 2DEGs is that the mobility decreases as
the density is reduced.~\cite{StormerAPL81} Thus at fixed $B$, the
discrepancy between the data and the solid line should decrease for
higher Landau levels $i$ where the density is higher, and we observe
this to be the case in Fig.~3. The slopes of the linear fits to
$\Delta \mu_{\rm T}$ versus $B$ for each $i$ are fairly consistent,
and slightly higher than expected (i.e., the solid line). This is
significant as the slope is directly related the effective mass
$m^{*}$. Averaging the slopes obtained for different $i$ values
gives $m^{*} = 0.042 m_{\rm e}$, which is qualitatively consistent
with the findings of Coleridge {\it et al.}, but $\sim 30 \%$ lower
than the value they obtain.~\cite{ColeridgeSS96} Towards addressing
this quantitative disagreement in $m^{*}$, we now briefly address
the main sources of error in our experiment. Averaging across the
three LLs, the error due to the linear fits is at most $11\%$, and
is dwarfed by a more dominant contribution due the dependence of the
disorder broadening on $n_{\rm T}$.~\cite{StormerAPL81} Each data
point for a given $i$ is obtained at a different $n_{\rm T}$, and
we estimate this could increase in the measured slope by up to
$25\%$, causing a decrease in the measured $m^{*}$ by a similar
amount. Finally, we note that the density dependence of $m^{*}$
should lead to non-linearities in the measured $\Delta \mu_{\rm T}$
versus $B$ data,~\cite{ColeridgeSS96} however these should be small
over the range studied, and are not evident in the data presented in
Fig.~3.

We conclude by discussing some potential improvements and
applications for our technique. The key limiting factor in our
device is the need to characterize and calibrate the sensor at each
field where data is obtained. This is due to the lack of a
back-gate, which prevents independent control of $n_{\rm B}$ in this
device. With a back-gate, we could establish a feedback mechanism
that uses the measured resistance to keep $n_{\rm B}$ constant. This
would greatly simplify sensor calibration and ensuring maximum
sensitivity over a greater $V_{\rm TG}$ range at arbitrary $B$. This
would also allow the measurements in Fig.~3 to be obtained at fixed
$n_{\rm T}$, overcoming the main source of error in measuring
$m^{*}$ with this device.

In summary, we have developed a sensitive electrometry system that
allows us to monitor the chemical potential of a 2D electron system
in the quantum Hall regime as its density is changed at fixed
magnetic field.  Our electrometer operates by exploiting the strong
capacitive coupling between two closely-spaced 2DESs in a double
quantum well heterostructure. As a demonstration of this device, we
have mapped the evolution of the Landau levels in a 2DES as a
function of magnetic field, and used this to measure the electron
effective mass, obtaining values that agree well with known
literature values.

This work was funded by Australian Research Council (ARC).  L.H.H.
acknowledges financial support from the UNSW and the CSIRO.

\end{document}